# Enhancing ordering dynamics in solvent-annealed block-copolymer films by lithographic hard masks supports

*Anja Stenbock-Fermor[†], Armin W. Knoll[‡], Alexander Böker[†,\*], Larisa Tsarkova[†,\*]*

[†]DWI – Leibniz-Institut für Interaktive Materialien, Forckenbeckstraße 50, 52056, Aachen Germany

[‡]IBM Research—Zurich, Säumerstrasse 4, CH-8803 Rüschlikon, Switzerland

KEYWORDS block copolymer films, lithographic hard masks, solvent vapor annealing, microdomain dynamics

ABSTRACT

We studied solvent-driven ordering dynamics of block copolymer films supported by a densely cross-linked polymer network designed as organic hard mask (HM) for lithographic fabrications. The ordering of microphase separated domains at low degrees of swelling corresponding to intermediate/strong segregation regimes was found to proceed significantly faster in films on a HM layer as compared to similar block copolymer films on silicon wafers. The ten-fold enhancement of the chain mobility was evident in the dynamics of morphological phase transitions and of related process of terrace-formation on a macroscale, as well as in the degree of long-range lateral order of nanostructures. The effect is independent of the chemical structure and on the volume composition (cylinder-/ lamella-forming) of the block copolymers. In-situ ellipsometric measurements of the swelling behavior revealed a cumulative increase in 1-3 vol.

% in solvent up-take by HM-block copolymer bilayer films, so that we suggest other than dilution effect reasons for the observed significant enhancement of the chain mobility in concentrated block copolymer solutions. Another beneficial effect of the HM-support is the suppression of the film dewetting which holds true even for low molecular weight homopolymer polystyrene films at high degrees of swelling. Apart from immediate technological impact in block copolymer-assisted nanolithography, our findings convey novel insight into effects of molecular architecture on polymer-solvent interactions.

INTRODUCTION

Thin supported polymer films are increasingly employed in many areas of nanofabrication. Chain dynamics and mobility in such layers is of significant fundamental interest as well as of essential technological importance because of their relevance to a large number of performance and processing issues,[1] e.g. in high-resolution lithography,[2] in the behavior of liquid crystal displays,[3] in organic photovoltaic[4-6] and in sensoring devices.[7] Also, a slow chain dynamics during annealing protocols of microphase separated structures in block copolymers films[8] is a limiting factor in developing cost-effective nanoscale patterning technologies.[9-10]

Thin films of block copolymers possess an intrinsic nano-structuring functionality which is provided by nanoscopic domains with tunable shapes, orientation and dimensions in the range of sub 10 nm up to 10s of nm. The knowledge acquired in the last decades on thermodynamic principles and dynamic mechanisms of the microphase separation[11-13], was sufficient to boost technology-oriented research on block copolymer films.[14-17] At the same time, these studies have established factors which complicate or crucially limit the application and performance of block copolymer films. One group of factors concerns the inherent thermodynamic features of the microphase separation.[18-20] Belonging to soft matter, block copolymers are characterized by low interfacial and surface energies, small differences in the free energy between different morphological states,[21] and relatively small incompatibility parameters between the blocks, which all lead to easy stabilization of meta-stable states and defects.[22-23] This makes the self-assembly of block copolymers in thin films extremely sensitive to external fields and to experimental uncertainties, such as tiny variations in the environmental temperature,[24] humidity,[25] solvent concentration[26-28] and quality,[29-32] surface energy at the substrate,[33-34] film preparation[35-37] and film thickness,[38] etc. While ambiguities in the experimental conditions are not always possible to eliminate and to control, the sensitivity of the microphase separation towards external fields, such as electric and shear fields, chemically and topographically patterned substrates, thermo- and solvent gradients, has turned to be advantageous for the nanofabrication of thin films, and in many research examples has been brought to a high level of control in guiding nanopatterns to perfectly ordered defect-free structures.[28, 39-44] Another group of factors which complicate the understanding and usage of block copolymer films is related to intrinsically slow dynamics of polymer chains[45] and to non-equilibrium aspects of film preparation and processing.[35-36]

In this respect, the solvent vapor annealing receives an increasing interest as a high-effective alternative to thermal processing of block copolymer films.[9] Recent examples of solvato-thermal annealing,[46] gradient solvent fields,[28, 47] of proximity injection approach[32] have demonstrated a reduction of processing times down to tens of seconds. However, up to now not much understanding on the relative role of the kinetics of solvent up take, equilibration in solvent vapors and quenching process on the resulting self-assembled structures has been reported.[9] On one side, recent studies suggest that ordered swollen states can have very fast dynamics that can lead to highly organized structures in short annealing times.[9,46] On the other side, the ordering times of less than a minute may imply that upon exposure to the solvent vapor the block copolymer undergoes order disorder transition (ODT), so that the removal of the solvent results in a dynamic quench from a disordered state. While such directional quench has been shown to be a promising route to control the orientation of microdomains,[48-50] this dynamic approach lacks the control over the block copolymer swelling level which is an essential issue for a targeted design of sizes and shapes of microdomains. In contrast, equilibration of structures under controlled conditions corresponding to intermediate or strong segregation regimes offers a number of advantages in fabrication of ordered patterns such as low line edge roughness of the resulting structures, fine control over non-bulk morphologies[38], and minimizing the risks of dewetting. Controlled solvent-vapor-annealing set up is based on a precise regulation of the solvent atmosphere in the annealing cell via solvent flow controllers and temperature control as well as on in-situ monitoring of the swollen film thickness, e.g. with optical reflectometry or in-situ ellipsometry (Figure 1a).[51] This procedure gives access to the value of the polymer volume fraction $\phi_p$ in a swollen film and thus allows for reproducible processing conditions below ODT. However, an insufficient chain dynamics is a current limitation of controlled annealing under reduced swelling. The work reported here discloses an approach toward a universal solution to this problem.

Apart from the technological benefits, controlled solvent vapor annealing provides a powerful scientific tool to fine-tune the microphase separation of block copolymers and thus to assess with a high level of sensitivity the whole spectra of polymer/solvent/substrate interactions as well as characteristic timescales of chains relaxations under confinement. Using this approach, Elbs and Krausch suggested a method to determine the Flory-Huggins interaction parameters in polymer thin films.[52] The experimental set up developed by Knoll et al[38] has been used to construct phase diagrams of surface structures in thin films of polystyrene-*b*-polybutadiene-*b*-polystyrene (SBS) triblock[51] and SB diblock copolymers,[53-54] as well as to quantitatively determine the energies of the morphological phase transitions.[21] Analysis of the phase behavior under controlled solvent vapor treatment allowed to gain important insights into the mechanisms of microphase separation of conjugated polymers,[5] and of microdomain reorientation and ordering under combined application of solvent annealing and of electric field.[55-58] Furthermore, confinement effects on the microphase separation and swelling of thin[24, 59] and thick[55, 37] block copolymer films have become evident in the thickness-dependent swelling as well as in heterogeneous swelling of block copolymer films.[24,60,61] In the last years similar climate-controlled solvent-vapor annealing devices have been adopted and successfully exploited by several research groups.[9] Additional valuable information on the microphase separated structures and their dimensions is gained when real-time grazing incidence small angle X-ray scattering (GISAXS) is utilized in combination with controlled solvent annealing.[62-63]

However, the presence of solvent molecules makes a block copolymer system more complicated for understanding as compared to polymer melts. Unlike theoretical studies, where system parameters such as surface fields, effective $\chi$ interaction parameters, and film thickness can be varied independently,[38] the above parameters in swollen films are affected simultaneously in a way which is not feasible to evaluate quantitatively. Despite its importance, the concentration dependence of $\chi$ parameters and of chain dynamics in swollen films has not been addressed systematically.[9] Also, the interaction of solvent molecules with the supporting substrate is an unclear issue.[64] Earlier studies have established that an increase in solvent concentration results in screening the interaction of block copolymers with silicon wafer supports.[38,51,61] In case of the substrates modified with soft organic layers such as grafted random copolymers or brushes,[50, 65] or photopatternable imaging layers,[66] the interactions of the modifying layer with the solvent and with the block copolymer chains can alter the swelling behavior and therefore the resulting microphase separation in block copolymer films.

Motivated by recent trends to incorporate patterning with block copolymers into complex lithographic fabrication schemes[66-67] such as thermal probe lithography[68] or soft graphoepitaxy with disposable photoresist patterns,[44] we used densely cross-linked networks from organic hard mask (HM) as supports for block copolymer films.[68] Since we employ solvent annealing as an advantageous processing step, we analyzed the effect of the underlying organic HM layer on the resulting microphase separated structures.

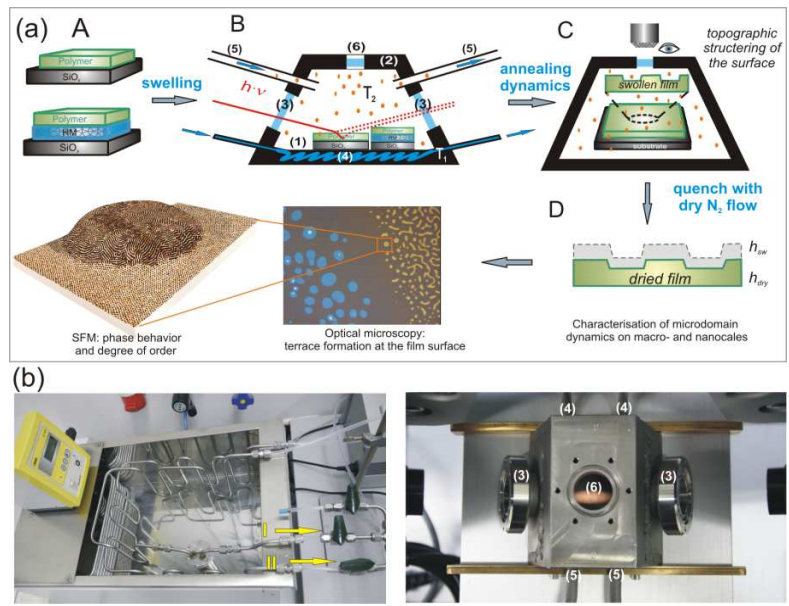

**Figure 1.** (a) A schematic of the experimental procedure. (**A**) Two types of samples used for combinatorial studies: block copolymer films supported by conventional silicon and by wafer with ~50 nm-thick HM coating (bilayer film). (**B**) Annealing chamber with (1) polymer samples; (2) sealed stainless steel walls; (3) glass windows, which are perpendicular to the incident light for in-situ ellipsometry; (4) temperature control of the substrate ($T_1$); (5) input and output for the temperature and flow controlled solvent vapor ($T_2$); (6) glass window for optical control of the terrace formation (**C**). (**D**) Characterization of quenched samples on macro- (optical microscopy) and nano- (SFM) scales for the topographical structuring of the free surface and for the phase behavior, respectively. (b) Photographs (left) of the thermostated solvent-vapor annealing system and (right) of the annealing chamber for in-situ ellipsometry measurements (top view). The solvent vapor atmosphere in the chamber is created and maintained by the combination of controlled flows of dry nitrogen (carrier gas) through the reservoir with a solvent (I) and of dry nitrogen (II). For notations see sketch in Figure 1a-B.

Here we report a significant improvement of the ordering dynamics of block copolymer films supported by lithographic organic hard mask (HM) as a model of a densely cross-linked polymer network. This phenomenon became evident upon comparing the microphase separation and the dynamic behavior of block copolymers in films supported by different substrates (conventional silicon wafers and HM layers) which have been simultaneously processed under controlled solvent vapor conditions (Figure 1a-A,B). For evaluation of the thickness of the swollen film $h_{sw}$ we employed both, direct in-situ ellipsometric measurements as well as indirect method based on the sensitive response of the microphase separation and of the chain mobility toward the solvent concentration in the swollen film. To do that, we chose cylinder-forming poly(styrene-*b*-butadiene) (SB) and lamella-forming polystyrene-*b*-poly(2-vinylpyridine) (SV) diblock copolymers which phase behavior has been earlier extensively studied under controlled solvent annealing protocols.[24,37,53,55] In particular, the phase transition to non-bulk perforated lamella (PL) morphology in SB films (see scanning force microscopy (SFM) image in Figure 1a-D) we considered as an "indicator" of certain intermediate segregation conditions in a swollen film, while the phase transition from vertically orientated to in-plane SV lamella, as well as the degree of the long-range order of microdomains were considered as a measure of the chain mobility. Moreover, both morphological transformations are accompanied by the development of surface relief structures (terraces) which allow following the annealing dynamics on a macroscale (Figure 1a C-D). The pronounced acceleration of block copolymer chain dynamics on HM-supports at reduced degrees of swelling can find immediate technological applications in block copolymer-assisted nanolithography.

EXPERIMENTAL SECTION

**Block copolymers.** Poly(styrene-*b*-butadiene) diblock copolymer (denoted as SB) with a total molecular weight of $M_n$ = 47.3 kg/mol and a polydispersity index of 1.03 was purchased from Polymer Source Inc. and used as received. The volume fraction of the PS block (26.1%) results in bulk morphology of hexagonally ordered PS cylinders with a characteristic spacing $d_0$ of 32.9 ± 0.3 nm and interlayer spacing of 3/2 ~ 27 nm.

Polystyrene-*b*-poly(2-vinylpyridine) diblock copolymer (denoted as SV) with a total molecular weight of $M_n$ = 99 kg/mol and a volume fraction of PS block of 0.56 was synthesized by sequential living anionic polymerization (polydispersity index 1.05). The characteristic lamellar spacing $L_0$ in bulk is about 47.2 nm.

**Substrates.** Polished and etched silicon wafers with a native silicon oxide top layer were purchased by CrysTec, cut in pieces and cleaned via ultrasonification in toluene for 10 min, followed by SnowJet and air-plasma (60 s) treatments directly before spin-coating.

Hardmask (HM)-supports have been prepared by spin coating the solution (HM8006-8) as purchased from JSR Micro at 5000 rpm. The coating is then cured on a hot plate at 225°C for 90s resulting in ~50 nm thick hard mask films.[68] The resulting HM layer represents a highly cross-linked glassy polymer network with high carbon content.

**Film preparation.** Polymer films were prepared from fresh, filtered solutions via spin coating on silicon substrates or HM supports. Chloroform and toluene (VWR) have been used without further purification to prepare SV and SB solutions in non selective solvents, respectively. The concentration of polymer solutions (0.5 - 1 wt%) and spinning rates during spincoating (2000 - 3000 rpm) have been chosen to gain block copolymer films with a thickness of ~ 50 nm. Films were additionally dried in vacuum at RT to remove the residual solvent. The thicknesses of spin-coated films have been measured with ellipsometry. In the case of block copolymer-HM bilayer films, the thickness of the block copolymer films was evaluated by subtracting the starting HM-coating thickness from the total bilayer thickness. When similar casting conditions have been used, the deviations in films thickness on silicon wafers and on HM-support were comparable with typical thickness deviations upon spin-coating procedure. To stabilize the polybutadiene (PB) against cross-linking, a small amount of stabilizer (2,6-die-*tert*-butyl-*p*-cresol) was added to solutions.

**Swelling experiments.** Solvent annealing of block copolymer films on two different types of substrates has been performed simultaneously in a custom-made climate-controlled liquid cell (Figure 1a). The setup allows for a precise control of the total flow through the chamber and of the solvent vapor concentration via flow controllers/flow meters (Brooks Instrument) at isothermal conditions (refrigerated circulator from Lauda). Another important controlled parameter is the swollen film thickness $h_{sw}$ which is constantly monitored via in-situ ellipsometry. The polymer fraction in a swollen film is then straightforward estimated as: $\phi$ =

$h_{dr}/h_{sw}$, where $h_{dr}$ is the thickness of a spin coated film and $h_{sw}$ of the respective film in a swollen state, correspondingly. The partial vapor pressure *p/p₀* of chloroform (a non-selective solvent for both studied block copolymers) has been adjusted according to the desired degree of swelling or polymer volume fraction by a combination of the flow of dry nitrogen with the flow of the saturated chloroform vapor created by guiding a flow of dry nitrogen through a solvent reservoir placed inside the thermostat. Flows are directed through the channel system made of Swagelog components.

Standard annealing settings has been used in comparative experiments: a total flow of 100 sccm, temperature was set to 20 °C for the solvent atmosphere and to 21.0°C for the substrates. The samples have been dried by purging a flow of dry nitrogen assuring fast vitrification of the structures[51] and then flushing the chamber with the nitrogen flow for 5 min before taking out the samples.

**Film characterization.** In-situ ellipsometry measurements have been made with Omt Imaging ellipsometer (mm30 series) at 70° incidence angle in a spectral range of 450-800 nm using VisuEl software 3.8. Analysis was made through fitting by Scout Software using a Cauchy model where block copolymer films and hard mask layers are modeled as a homogeneous material.

As spin-coated and annealed block copolymer films has been characterized with optical microscopy (Keyence digital optical microscope, VHX2000) for surface relief structures. Then microphase separation behavior has been analyzed with scanning force microscopy (SFM) measurements in intermittent mode (Bruker Dimension Icon) using Nanoscope 8.10 software and OTESPA tips (spring constant 12-103 N/m, resonant frequency 278-357 kHz) under ambient conditions.

RESULTS AND DISCUSSION

Block copolymer films have been spin coated from respective solutions either on plasma-cleaned silicon wafers or on HM layer, followed by a drying in a vacuum oven to remove the residual solvent (an important step before measuring the starting thickness in a dry state). In each comparative solvent-treatment experiment, similarly prepared block copolymer films have been placed simultaneously in the annealing chamber, and the swollen thickness of silicon wafer-supported films has been continuously measured with in-situ ellipsometry to monitor the stability of the annealing conditions (Figure 1a-A,B). After a fast quench with a flow of dry nitrogen the microphase separation behavior was analyzed by SFM.

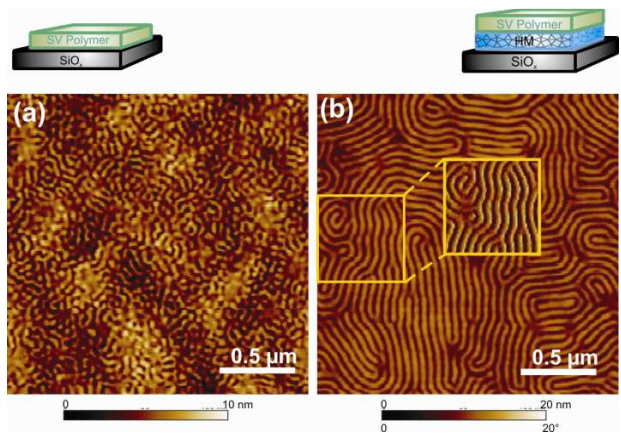

**Figure 2.** SFM topography images of SV films on silica-substrate (a) and on HM-support (b) after annealing at 20% of $p/p_0$ ($\phi_p$ of 0.76±0.02) for 1 h. The pattern of the up standing lamella is preformed during spin coating since the dry thickness of ~50 nm (~$L_0$) is below a critical thickness of 1.5 $L_0$ to form in-plane oriented lamella. Inset in (b) is the respective phase image of the highlighted area in the topography image.

By testing block copolymers with varied chemical composition and volume fraction of the blocks and films with varied thickness we systematically observed a higher degree of long range order of microdomains in HM-supported samples as compared to similar films on silicon wafers (Figure 2). The faster ordering dynamics might be associated with a decrease of block copolymer viscosity due to a higher solvent concentration, i.e. higher degree of swelling in HM-supported films. To clarify this scenario, we performed controlled solvent vapor annealing experiments under standard conditions regarding the value of the total continuous flow $p_0$ through the chamber and the temperatures of the solvent vapor and of the substrate (Figure 1a,b). The partial solvent vapor pressure $p/p_0$ has been adjusted by mixing controlled flows of dry nitrogen and of carrier gas (dry nitrogen) saturated with the vapor of chloroform $(p)$, a non-selective solvent for studied here block copolymers.

Figure 3a presents kinetics of swelling upon step-wise increase of $p/p_0$ of SV film supported by silicon wafer. The saturation of the SV film with chloroform vapor, i.e. the constant thickness at each swelling step is achieved within about ten minutes. As seen in Figure 3b, swelling of a HM layer (of similar thickness as the SV film) is notably less and slower, so that saturation cannot be reached even on a time scale of several hours (Figure 1S, Supporting information).

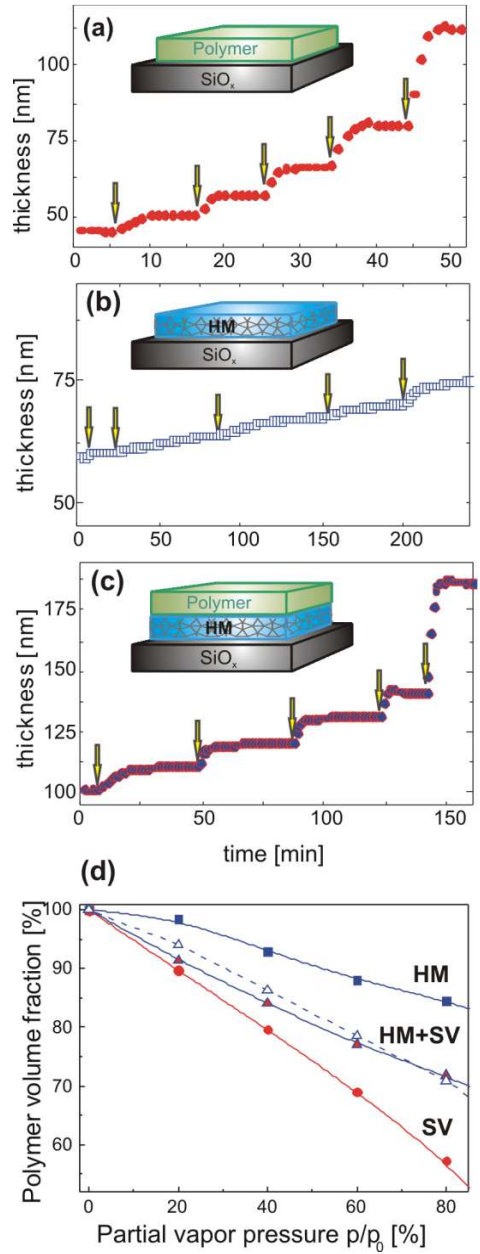

**Figure 3.** Kinetics of swelling upon step-wise increase of the partial vapor pressure $p/p_0$ of chloroform (as indicated by arrows from left to right) to 20, 40, 60, 80 and 100% of (a) SV film on a silicon wafer, (b) HM-layer on a silicon wafer, and (c) SV film on HM-support. (d) Comparison of the swelling behavior evaluated from the corresponding swelling curves in (a-c) and expressed as a polymer volume fraction $\phi_p$ versus $p/p_0$ for HM-layer (squares), for SV film on silicon substrate (circles) and SV film on HM-support (filled triangles). Empty triangles indicate calculated values assuming a summation of the degrees of swelling of respective layers.

The swelling kinetics of a SV film on a HM support (a bilayer film) is displayed in Figure 3c. The starting dry thickness of the bilayer is similar to the sum of the HM and SV film thickness on silicon wafers under identical spin coating conditions. Thus we conclude that the HM support does not significantly affect the resulting thickness of spin coated block copolymer films.

As seen in Figure 3c, the time to saturation at each swelling step for a SV-HM bilayer is approximately twice larger as that of SV film on a silicon wafer. The swelling kinetic of the SV-HM bilayer film is a combination of the solvent uptake by both layers which can have a mutual effect on each other. In particular, we anticipate that the swelling of the HM layer proceeds faster compared to the pure HM film due to the direct contact with the swollen SV film. We note that the differences in the swelling behavior of the individual layers could not be resolved by fitting the ellipsometric data with the Cauchy model due to the similarity refractive indexes of both materials. An advanced modeling might solve this problem.[69]

Figure 3d presents a summary of the swelling data which is displayed as polymer volume fraction $\phi_p$ in the film ($\phi_p = h_{dr}/h_{sw}$) versus $p/p_0$. For comparison the extrapolated $\phi_p$ value for HM-SV bilayer is shown, given by the sum of $\phi_p$ of the respective individual layers (with 50:50 volume ratio). The deviation is on the order of 1-3%, which is comparable to the accuracy of the degree of swelling under controlled experimental conditions. The swelling data for the films with a higher dry thickness of ~112 nm is presented in Figure 2S (Supporting information) and reveals qualitatively similar behavior. We conclude that the influence of an underlying HM layer on the solvent up-take of a block copolymer film is not significant.

To quantify the non-trivial effect of the enhanced ordering dynamics without a conclusive increase in solvent uptake, we used an indirect method based on the sensitive response of the microphase separation and of the chain dynamics toward the solvent concentration in the film. These responses can be accessed with the optical microscopy and with SFM on macro- and nano-scales. In particular, when processing SV films we found that as long as the averaged film thickness is below a critical value of $h_{crit}$ ~ 75 nm, the lamella is oriented perpendicular to the film plane. When the swollen thickness is adjusted to ~ $h_{crit}$, the lamella starts to reorient parallel to the film plane and forms terraces. This dynamic process can be observed by optical microscopy as schematically depicted in Figure 4a.

As shown earlier, the morphological behavior of block copolymers containing PS and PVP compartments in thin films on silicon wafers is guided by the preferential wetting of the substrate by the PVP block, while the PS block is segregated to the free surface due to its lower surface tension resulting in asymmetric wetting conditions.[55] The equilibrium thickness of an in-plane lamella is proportional to (0.5+n) $L_0$, which amounts to approximately 75 nm for 1.5$L_0$-thick SV films, consistent with the observed $h_{crit}$.

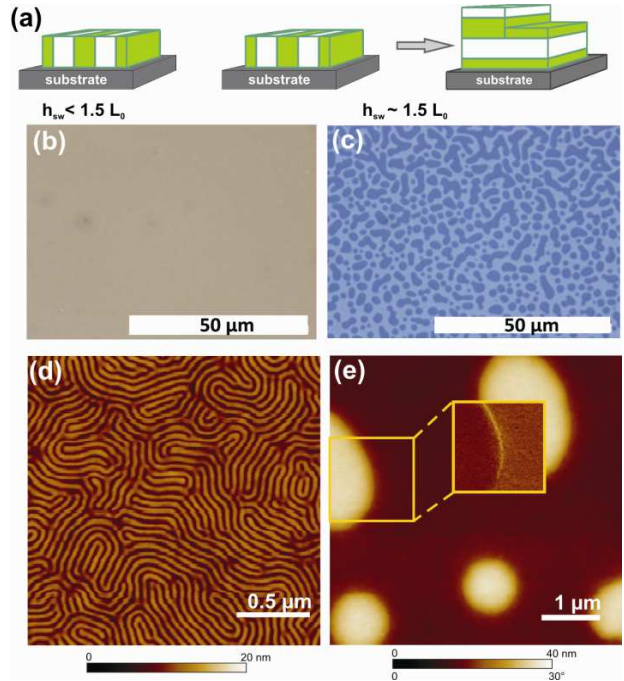

**Figure 4.** a) A sketch of the phase transition from perpendicular to in-plane oriented lamella with asymmetric wetting conditions in SV films when the thickness in swollen state $h_{sw}$ exceeds a critical value of 1.5 $L_0$. Optical microscopy (b,c) and SFM topography (d,e) images of SV films on silica-substrate (b,d) and on HM-support (c,e) after annealing at 80% ($\phi_p$ of 0.66±0.02) for 30 min under standard thermal and flow conditions. Inset in (e) is the corresponding phase image of the highlighted area in the topography image. The dry thickness of the films is ~50 nm (~$L_0$) similar to that of the films shown in Figure 2.

Figure 4 b,c presents optical microscopy images of SV films with the starting thickness ~$L_0$. Both films have been exposed to 80% of $p/p_o$ in order to reach $h_{crit}$. Upon 30 min of equilibration, the SV film on silicon wafer is characterized by a uniform color indicating a smooth surface topography which is preserved after spin-coating and indicates very limited chain mobility in the film (Figure 4b). In contrast, HM-supported SV film, processed under identical annealing conditions, exhibits a pattern of two distinct colors which are characteristic of a well developed terrace formation process (Figure 4b). Accordingly, SFM topography images in Figure 4d,e display a striped pattern of upstanding lamella on silicon wafer, and coexisting terraces of featureless in-plane oriented lamella on HM support (Figure 4e). This comparison of the time dependence of the lamella reorientation indicates a highly accelerated dynamics of the morphological transition in the case of HM-supported block copolymer films.

Obviously, the long range order of lamella stripes on silicon wafer upon annealing at 80% of *p/p$_o$* (Figure 4d) is much more advanced as compared to the degree of order in the film annealed at lower swelling (Figure 2a). However the period of 30 min was not sufficient to initiate the transition to thermodynamically favorable in-plane lamella morphology. To follow further the time-scale of the structure development, we performed a long-term annealing of the SV film on silicon wafer under similar conditions. Figure 5 presents the optical image (a) and SFM image (b) of the SV film upon 9 h of equilibration. Both images indicate almost complete transition to in-plane oriented lamella, confirming the similarity of the wetting conditions in SV films on silicon wafers and on HM support.

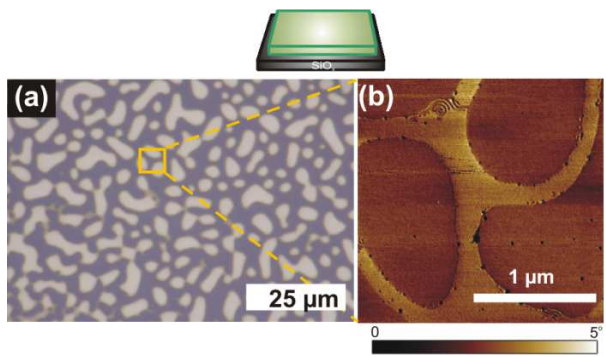

**Figure 5.** Optical microscopy (a) and SFM phase (b) images of 50 nm-thick SV film on silicon wafer annealed for 9 h under standard conditions allowing maintaining ϕ$_p$ at a constant level of 0.65±0.02.

We may use the observed timescales on the two different substrates to quantify the enhanced chain diffusion in swollen films supported by HM layer. Using Einstein`s relation a = (Dt)$^{1/2}$ with a~L$_0$~50 nm, and respective t$_{si}$ ~ 9 h and t$_{HM}$ ~ 0.5 h results in one order of magnitude larger diffusion constant in HM-supported films.

Another important aspect of microphase separation is the sensitivity of the structures to the χ-parameter which can be fine-tuned by the solvent concentration in the film. We used this responsive behavior as a "sensor" of the segregation power in simultaneously processed SB films on silicon wafer and on HM support. The advantage of this test as compared to in-situ ellipsometric studies (Figure 3) is the possibility to avoid uncontrolled environmental fluctuations which can occur in separate annealing runs. Also this analysis allows to evaluate a sole contribution of the block copolymer layer to the total swelling of the SB-HM bilayer film.

Shown in Figure 6 are SFM images of solvent-annealed films of cylinder-forming SB block copolymer. Initial dry thickness of ~50 nm corresponds to one and a half layers of structures. The SB film on silicon wafer shows a flat surface topography (Figure 6a) with the phase behavior representing a disordered mixture of striped pattern (PS cylindrical domains in soft PB matrix), small patches of black dots (PL phase) and a high concentration of unfavorable defects such as white dots or horse-shoe defects[70] (Figure 6c). In contrast, images of HM-supported SB film reveal terraces with two distinct thicknesses (Figure 6b) and spatially-separated phases of PL and of in-plane oriented cylinders ($C_\parallel$). The higher degree of long-range order, as well as the terrace formation in HM-supported films convincingly demonstrates a faster ordering dynamics as compared to SB film on silicon wafer which shows very limited chain mobility.

The deviation from the bulk cylinder morphology can be explained using an earlier established phase diagram of surface structures in swollen SB films as a function of the solvent content (chloroform) in the film.[53-54,61] According to the phase diagram (Figure 6e), the PL phase and the $C_\parallel$ phase develop in the first and in the second terrace, respectively, upon annealing at reduced solvent vapor atmosphere $p/p_o$ ~ 50-60% (corresponding to $\phi_p$ ~ 0.7÷0.8). Here an "indicator" of the intermediate segregation conditions is the PL phase – a deviation from the cylinder bulk morphology – which vanishes when the solvent concentration in the film becomes higher than 30%. Since in SB film on HM-support the PL phase is a stable morphology, we conclude that the swollen SB film on HM layer is within the intermediate segregation regime.[54]

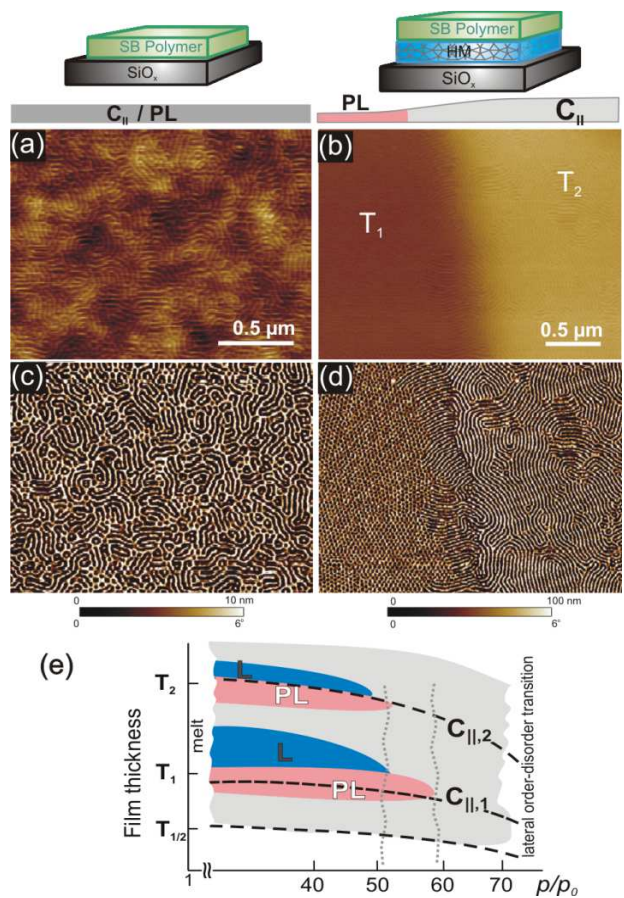

**Figure 6.** SFM topography (a,b) and phase (c,d) images of the microphase separated structures in ~50 nm-thick SB films on silica wafer (a, c) and on HM-support (b, d) after annealing for 1 h in chloroform vapor under standard thermal and flow conditions at 55% of $p/p_0$ ($\phi_p \sim 0.8$). The corresponding sketches of the film topography above the images indicate the development of the coexisting $C_\parallel$ and PL phases on silica substrate. On HM-support the PL phase and the $C_\parallel$ phase develop in terrace $T_1$ and $T_2$, respectively. (e) A schematic of the phase diagram of the surface structures in solvent-annealed SB films, reported in Reference[54]. Dashed vertical lines mark the area which corresponds to the annealing conditions as in Figures 6 and 7.

Figure 7 a,b presents optical microscopy images of SB films shown in Figure 6. Well-developed terraces on HM-support versus flat surface topography of SB film on silicon wafer are a clear sign of enhanced dynamics. Achieving the same degree of topographical structuring of the free surface under comparable annealing conditions on silicon wafers typically takes several hours.[61,70] Moreover, optical microscopy images Figure 7 c,d, demonstrate that already short-term processing for 10 min of SB films under reduced solvent atmosphere results in much more advanced terracing process as compared to silicon wafers-supported films. The SFM height and phase images of the respective films are shown in Figure 3S (Supporting information).

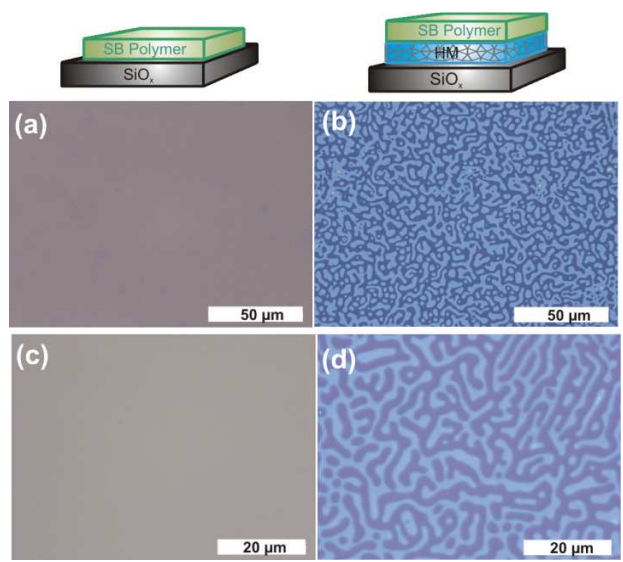

**Figure 7.** Optical microscopy images of annealed SB films on silica-support (a, c) and on HM-support (b, d). Annealing times are 1 hour (a,b) and 10 min (c,d). The annealing conditions for films in (a,b) are described in Figure 6. Films shown in (c,d) with the dry thicknesses of 45 nm have been annealed for 10 min under standard thermal and flow conditions at 60% of $p/p_0$ ($\phi_p$ ~0.7).

To our current understanding, a significant improvement of the solvent-driven ordering dynamics of block copolymer films on HM supports can be attributed to the reduced interactions at the block copolymer / HM interface. On the other side, weak interactions with the substrate during annealing procedures diminish the stability of the polymer films towards dewetting, a concurrent to microdomain ordering dynamic process. Generally in our annealing experiments, HM-supported films were stable towards dewetting while at the same time revealing a high ordering dynamics.

Since thin PS homopolymer films, especially of low molecular weight PS, are known to show high dewetting instability upon annealing procedures, we performed comparative swelling studies at 100% of $_{p/p0}$ of PS films on silicon wafer and on HM-support. Figure 8 displays optical microscopy images of as spin-coated and solvent-processed PS films which reveal a clear retardation of dewetting of HM-supported film (Figure 8d) compared to completely dewetted PS film on silicon wafer (Figure 8c). Interestingly, qualitatively similar effect of suppression of dewetting on HM-supports was observed for thermally annealed (at 120°C) PS films.

This predictable observation can be presumably explained by partial interdigitation of linear block copolymer chains with the polymer network-structure of the HM support. Our current on-going work aims at determining solvent-dependence of the glass transition of the HM layer and its effect on the dynamic behavior of polymer films. We note, that the opposite effects of the HM-support on two dynamic processes associated with the processing of polymer films (suppression of a destructive dewetting and enhancements of the equilibration dynamics) is a non-trivial result with promising benefits in block copolymer-related technological fabrications.

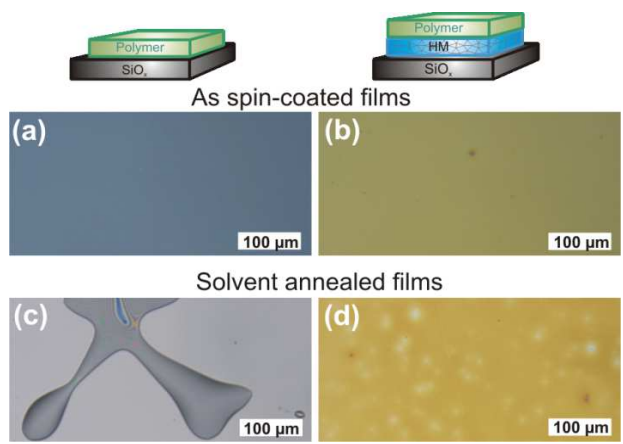

**Figure 8.** Optical microscopy images of ~140 nm-thick PS homopolymer ($M_n$=5.6 kg/mol) films on (a,c) silicon wafer and (b,d) on a HM-support after spin coating (a,b) and after 10 min of annealing at 100% of *p/p₀* under standard flow and temperature conditions.

CONCLUSIONS

We report here a dual beneficial effect of a supporting layer from a densely cross-linked organic hard mask (HM) on the ordering dynamics and stability of solvent-processed block copolymer films. A series of combinatorial experiments on block copolymers with varied chemical composition and volume fraction of the blocks and with varied film thickness revealed a pronounced enhancement of the microdomain ordering in HM-supported samples as compared to similar films on silicon wafers. Since ellipsometric measurements did not reveal a definite increase in solvent up-take by HM-supported films, the dilution effect cannot account for the enhanced ordering dynamics. Hence the properties of the swollen HM-support are envisaged to be decisive in accelerating block copolymer chain dynamics at reduced degrees of swelling.

To confirm the non-trivial effect of the enhanced ordering dynamics without a conclusive change in the degree swelling, we used an indirect method based on the "sensitivity" of the microphase separation and of the chain mobility toward the solvent concentration in a film. Phase transitions to non-bulk PL morphologies in cylinder-forming block copolymers have been considered as "indicators" of particular segregation conditions, while phase transition from in-plane to vertically orientated lamella and long-range order of microdomains were considered as a measure of the chain mobility. Importantly, enhanced ordering was achieved even at a reduced degree of controlled swelling corresponding to an intermediate to strong segregation regime, when similar films on conventional substrate show very limited mobility.

Apart from direct practical applications in block copolymer nanolithography, in particular, for the systems with large $\chi$ interaction parameter with high molecular weights, our findings raise the fundamental issues concerning the effect of confinement and of polymer architecture on polymer-solvent interactions.

AUTHOR INFORMATION

**Corresponding Author**

*Authors to whom correspondence should be addressed. E-mail: boker@dwi.rwth-aachen.de; tsarkova@dwi.rwth-aachen.de

**Author Contributions**

The manuscript was written through contributions of all authors. All authors have given approval to the final version of the manuscript.

ACKNOWLEDGMENT

K. Schmidt is acknowledged for the synthesis of SV polymer. AK acknowledges the financial support from the Starting Grant 307079 of the European Research Council (ERC).

*Supporting Information Available.* Extensive figures 1S-3S. This material is available free of charge via the Internet at http://pubs.acs.org/.

REFERENCES

1. Ediger, M. D.; Forrest, J. A. *Macromolecules* **2013**, *47 (2)*, 471–478.

2. Saavedra, H. M.; Müllen, T. J.; Zhang, P.; Dewey, D. C.; Claridge, S. A.; Weiss, P. S. *Rep. Prog. Phys.* **2010**, *73*, 036501.

3. Gibbons, W. M.; Shannon, P. J.; Sun, S.-T.; Swetlin, B. J. *Nature* **1991**, *351*, 49-50.

4. Schmidt-Mende, L.; Fechtenkötter, A.; Müllen, K.; Moons, E.; Friend, R. H.; MacKenzie, J. D. *Science* **2001**, *293*, 1119-1122.

5. Hüttner, S.; Sommer, M.; Chiche, A.; Krausch, G.; Steiner, U.; Thelakkat, M. *Soft Matter* **2009**, *5*, 4206-4211.

6. Johnson, K.; Huang, Y.-S.; Hüttner, S.; Sommer, M.; Brinkmann, M.; Mulherin, R.; Niedzialek, D.; Beljonne, D.; Clark, J.; Huck, W. T. S.; Friend, R. H. *J. Am. Chem. Soc.* **2013**, *135*, 5074-5083.

7. Späth, K.; Gauglitz, G. *Mater. Sci. Eng.,* **1998**, *C5*, 187-191.

8. Fredrickson, G. H.; Bates, F. S. *Annu. Rev. Mater. Sci.* **1996**, *26*, 501-550.

9. Sinturel, C.; Vayer, M.; Morris, M.; Hillmyer, M. A. *Macromolecules* **2013**, *46*, 5399-5415.

10. Albert, J. N. L.; Epps, T. H., III *Mater. Today* **2010**, *13*, 24-33.

11. Hamley, I. W., The Physics of Block Copolymers. *Oxford University Press: Oxford* **1998**.

12. Fasolka, M. J.; Mayes, A. M. *Annu. Rev. Mater. Res.* **2001**, *31*, 323-355.

13. Lodge, T. P. *Macromol. Chem. Physic.* **2003**, *204*, 265-273.

14. Park, C.; Yoon, J.; Thomas, E. L. *Polymer* **2003**, *44*, 6725-6760.

15. Stoykovich, M. P.; Müller, M.; Kim, S. O.; Solak, H. H.; Edwards, E. W.; de Pablo, J. J.; Nealey, P. F. *Science* **2005**, *308*, 1442-1446.

16. Kim, D. H.; Lau, K. H. A.; Robertson, J. W. F.; Lee, O.-J.; Jeong, U.; Lee, J. I.; Hawker, C. J.; Russell, T. P.; Kim, J. K.; Knoll, W. *Adv. Mater.* **2005,** *17*, 2442-2446.

17. Thurn-Albrecht, T.; Schotter, J.; Kastle, G. A.; Emley, N.; Shibauchi, T.; Krusin-Elbaum, L.; Guarini, K.; Black, C. T.; Tuominen, M. T.; Russell, T. P. *Science* **2000**, *290*, 2126-2129.

18. Matsen, M. W.; Bates, F. S. *Macromolecules* **1996**, *29*, 1091-1098.

19. Matsen, M. W. *J. Chem. Phys.* **1997,** *106*, 7781-7791.

20. Farrell, R. A.; Fitzgerald, T. G.; Borah, D.; Holmes, J. D.; Morris, M. A. *Int. J. Mol. Sci.* **2009,** *10*, 3671-3712.

21. Knoll, A.; Lyakhova, K. S.; Horvat, A.; Krausch, G.; Sevink, G. J. A.; Zvelindovsky, A. V.; Magerle, R. *Nat. Mater.* **2004,** *3*, 886-891.

22. Harrison, C.; Adamson, D. H.; Cheng, Z.; Sebastian, J. M.; Sethuraman, S.; Huse, D. A.; Register, R. A.; Chaikin, P. M. *Science* **2000,** *290*, 1558-1561.

23. Hammond, M. R.; Sides, S. W.; Fredrickson, G. H.; Kramer, E. J.; Ruokolainen, J.; Hahn, S. F. *Macromolecules* **2003,** *36*, 8712-8716.

24. Zettl, U.; Knoll, A.; Tsarkova, L. *Langmuir* **2010,** *26*, 6610-6617.

25. Park, S. C.; Kim, B. J.; Hawker, C. J.; Kramer, E. J.; Bang, J.; Ha, J. S. *Macromolecules* **2007,** *40*, 8119-8124.

26. Kim, G.; Libera, M. *Macromolecules* **1998,** *31*, 2569-2577.

27. Kim, H.-C.; Russell, T. P. **2001,** *39*, 663-668.

28. Kim, S. H.; Misner, M. J.; Xu, T.; Kimura, M.; Russell, T. P. *Adv. Mater.* **2004,** *16*, 226-231.

29. Jung, Y. S.; Ross, C. A. *Adv.Mater.* **2009,** *21*, 2540.

30. Bosworth, J. K.; Black, C. T.; Ober, C. K. *ACS Nano* **2009,** *3*, 1761-1766.

31. Hagaman, D.; Enright, T. P.; Sidorenko, A. *Macromolecules* **2012,** *45*, 275-282.

32. Jeong, J. W.; Hur, Y. H.; Kim, H.-j.; Kim, J. M.; Park, W. I.; Kim, M. J.; Kim, B. J.; Jung, Y. S. *ACS Nano* **2013,** *7*, 6747-6757.

33. Mansky, P.; Russell, T. P.; Hawker, C. J.; Mays, J.; Cook, D. C.; Satija, S. K. *Phys. Rev. Lett.* **1997,** *79*, 237-240.

34. Peters, R. D.; Yang, X. M.; Kim, T. K.; Sohn, B. H.; Nealey, P. F. *Langmuir* **2000,** *16*, 4625-4631.

35. Reiter, G.; Hamieh, M.; Damman, P.; Sclavons, S.; Gabriele, S.; Vilmin, T.; Raphael, E. *Nat. Mater.* **2005,** *4*, 754-758.

36. Damman, P.; Gabriele, S.; Coppee, S.; Desprez, S.; Villers, D.; Vilmin, T.; Raphael, E.; Hamieh, M.; Al Akhrass, S.; Reiter, G. *Phys. Rev. Lett.* **2007,** *99*, 036101.

37. Gensel, J.; Liedel, C.; Schoberth, H. G.; Tsarkova, L. *Soft Matter* **2009,** *5*, 2534-2537.

38. Knoll, A.; Horvat, A.; Lyakhova, K. S.; Krausch, G.; Sevink, G. J. A.; Zvelindovsky, A. V.; Magerle, R. *Phys. Rev. Lett.* **2002,** *89*, 035501.

39. Segalman, R. A. *Mater. Sci. Eng.* **2005,** *R48*, 191-226.

40. Darling, S. B. *Prog. Polym. Sci.* **2007,** *32*, 1152-1204.

41. Segalman, R. A.; Yokoyama, H.; Kramer, E. J. *Adv. Mater.* **2001,** *13*, 1152-1155.

42. Bita, I.; Yang, J. K. W.; Jung, Y. S.; Ross, C. A.; Thomas, E. L.; Berggren, K. K. *Science* **2008,** *321*, 939-943.

43. Ruiz, R.; Kang, H. M.; Detcheverry, F. A.; Dobisz, E.; Kercher, D. S.; Albrecht, T. R.; de Pablo, J. J.; Nealey, P. F. *Science* **2008,** *321*, 936-939.

44. Jeong, S.-J.; Moon, H.-S.; Kim, B. H.; Kim, J. Y.; Yu, J.; Lee, S.; Lee, M. G.; Choi, H.; Kim, S. O. *ACS Nano* **2010,** *4*, 5181-5186.

45. Cavicchi, K. A.; Lodge, T. P. **2004,** *37*, 6004-6012.

46. Gotrik, K. W.; Ross, C. A. *Nano Lett.* **2013,** *13*, 5117-5122.

47. Yoon, J.; Lee, W.; Thomas, E. L. *Adv. Mater.* **2006,** *18*, 2691.

48. Kim, S. H.; Misner, M. J.; Xu, T.; Kimura, M.; Russell, T. P. *Adv. Mater.* **2004,** *16*, 226.

49. Phillip, W. A.; Hillmyer, M. A.; Cussler, E. L. *Macromolecules* **2010,** *43*, 7763-7770.

50. Jeong, J. W.; Park, W. I.; Kim, M.-J.; Ross, C. A.; Jung, Y. S. *Nano Lett.* **2011,** *11*, 4095-4101.

51. Knoll, A.; Magerle, R.; Krausch, G. *J. Chem. Phys.* **2004,** *120*, 1105-1116.

52. Elbs, H.; Krausch, G. *Polymer* **2004,** *45*, 7935-7942.

53. Tsarkova, L. In *Nanostructured Soft Matter: Experiment, Theory, Simulation and Perspectives*, Zvelindovsky, A. V., Ed.; Springer: Heidelberg, **2007**.

54. Tsarkova, L.; Sevink, G. J. A.; Krausch, G. *Adv. Polym. Sci.* **2010,** *227*, 33-73.

55. Olszowka, V.; Tsarkova, L.; Böker, A. *Soft Matter* **2009,** *5*, 812-819.

56. Olszowka, V.; Hund, M.; Kuntermann, V.; Scherdel, S.; Tsarkova, L.; Böker, A. *ACS Nano* **2009,** *3*, 1091-1096.

57. Liedel, C.; Hund, M.; Olszowka, V.; Boeker, A. *Soft Matter* **2012,** *8*, 995-1002.

58. Liedel, C.; Pester, C. W.; Ruppel, M.; Lewin, C.; Pavan, M. J.; Urban, V. S.; Shenhar, R.; Boesecke, P.; Böker, A, *ACS Macro Lett.* **2013,** *2*, 53-58.

59. Tsarkova, L. In *Trends in Colloid and Interface Science XXIII*, Bucak, S., Ed. Springer-Verlag: Berlin, **2010**; Vol. 137.

60. Papadakis, C. M.; Di, Z.; Posselt, D.; Smilgies, D.-M. *Langmuir* **2008**, *24*, 13815-13818.

61. Tsarkova, L. *Macromolecules* **2012,** *45*, 7985-7994.

62. Di, Z.; Posselt, D.; Smilgies, D.-M.; Li, R.; Rauscher, M.; Potemkin, I. I.; Papadakis, C. M. *Macromolecules* **2012,** *45*, 5185-5195.

63. Paik, M. Y.; Bosworth, J. K.; Smilges, D.-M.; Schwartz, E. L.; Andre, X.; Ober, C. K. *Macromolecules* **2010,** *43*, 4253-4260.

64. Vogt, B. D.; Soles, C. L.; Jones, R. L.; Wang, C.-Y.; Lin, E. K.; Wu, W.-l.; Satija, S. K.; Goldfarb, D. L.; Angelopoulos, M. *Langmuir* **2004,** *20*, 5285-5290.

65. Mansky, P.; Russell, T. P.; Hawker, C. J.; Pitsikalis, M.; Mays, J. *Macromolecules* **1997,** *30*, 6810-6813.

66. Han, E.; In, I.; Park, S.-M.; La, Y.-H.; Wang, Y.; Nealey, P. F.; Gopalan, P. *Adv. Mater.* **2007,** *19*, 4448-4452.

67. Cheng, J. Y.; Sanders, D. P.; Truong, H. D.; Harrer, S.; Friz, A.; Holmes, S.; Colburn, M.; Hinsberg, W. D. *ACS Nano* **2004***, 4*, 4815-4823.

68. Cheong, L. L.; Paul, P.; Holzner, F.; Despont, M.; Coady, D. J.; Hedrick, J. L.; Allen, R.; Knoll, A. W.; Duerig, U. *Nano Lett.* **2013,** *13*, 4485-4491.

69. Ogieglo, W.; Wormeester, H.; Wessling, M.; Benes, N. E. *Macromol. Chem. Physic.* **2013,** *214*, 2480-2488.

70. Horvat, A.; Sevink, G. J. A.; Zvelindovsky, A. V.; Krekhov, A.; Tsarkova, L. *ACS Nano* **2008,** *2*, 1143-1152.